\documentstyle[12pt,aasms4]{article}

\received{July 9, 1997}
\accepted{December 5, 1997}

\newcommand{\kmps}{{\rm \,km\,s$^{-1}\;$}}            
\newcommand{\degree}{$^\circ\;$}                    
\newcommand{\Vlsr}{{\rm \,V$_{lsr}\,$}}               

\slugcomment{(ApJL in press ;version 1.6, 11/17/97)}

\lefthead{Izumiura et al.}
\righthead{SiO Masers at the Galactic Center}

\begin{document}

\title{SiO Maser Forest at the Galactic Center
}

\author{Hideyuki Izumiura}
\affil{Okayama Astrophysical Observatory, 
National Astronomical Observatory, \\
Kamogata, Asakuchi, Okayama 719-02, Japan\\
e-mail: izumiura@oao.nao.ac.jp}

\author{Shuji Deguchi}
\affil{Nobeyama Radio Observatory, National Astronomical Observatory,\\
          Minamimaki, Minamisaku, Nagano 384-13, Japan\\
          e-mail: deguchi@nro.nao.ac.jp}

\and 
\author{Takahiro Fujii}
\affil{Department of Astronomy, Faculty of Science, the University of Tokyo\\
       Bunkyo-ku, Tokyo 110, Japan\\
       e-mail: fujii@astron.s.u-tokyo.ac.jp}
\begin{abstract}

   A moderately deep survey of stellar maser sources
toward the Galactic center has been made in the SiO J=1-0 v=1 
and v=2 transitions ($\sim 43$ GHz) with the Nobeyama 45-m telescope 
of a beam size of about 40$''$.  At the Galactic center
(Sgr A*), 7 (4 new) 
SiO maser sources were detected in one beam. 
In order to investigate the spatial distribution of SiO maser sources,
we have also observed the offset positions 
by $\pm$ 40$''$ in Galactic longitude from the Galactic center. 
In total, 7 (4 new) SiO maser
sources were detected at the offset positions. 
Taking into account the shorter integration time 
at the offset positions, 
we find that the surface number density of SiO maser sources 
is nearly constant at the Galactic center. The number density
of SiO maser sources is found to be an order of magnitude higher 
than the density of OH 1612 MHz objects.
A radial-velocity gradient in Galactic longitude
was not detected. These facts indicate that the SiO maser 
sources seen toward the Galactic center are mostly associated 
with the stellar population of a Galactic stellar nuclear disk 
of more than a few arc minute radius.

\end{abstract}


\section{Introduction}

   It is known that a cluster of evolved stars surrounds 
the Galactic center 
(\cite{blu96}; \cite{mor96})\footnote{We refer to the dynamical center 
of our Galaxy as the radio continuum source, 
Sgr A*, in this paper.}. 
These evolved stars very often exhibit molecular line masers 
(\cite{win96}) and have been used for studying the structure 
of the Galactic center disk. OH 1612 MHz and SiO 43 GHz 
maser lines have been used to obtain radial velocities for sources
in the Galactic center region (\cite{lin91}; \cite{lin92a}), 
for the Galactic bulge IRAS sources (\cite{izu95}), and 
for sources toward the Sgr B2 molecular cloud (\cite{shi97}).  
Maser sources very close to the Galactic center are useful 
to establish a concordance between radio 
and infrared position-reference frames  (\cite{men97}),
identifying the Sgr A* in the infrared (\cite{eck93}),
and studying proper motions (\cite{eck97}). 
With the Very Large Array (VLA), 
sensitive surveys of maser sources near Sgr A*
have been made in the OH 1612 MHz line (e.g., \cite{sjo97}). 
A VLA survey in  SiO/H$_{2}$O lines (\cite{men97}) was made with 
velocity resolution of 2.7 \kmps with 32 channels,
resulting in two new SiO maser sources
associated with known infrared 
objects very near Sgr A*. 
The detection limit
of this VLA survey was at the level of $\sim$0.05 Jy at 43 GHz.

In this paper, we report an attempt of surveying  
SiO maser sources near the Galactic center 
with the Nobeyama 45-m telescope. 
We have detected 7 SiO sources at the Galactic center. 
In order to investigate the source density near the Galactic center 
we have also observed at the two offset positions by $\pm 40 ''$   
in Galactic longitude, resulting in detections of 
similar number of sources. 

\section{Observations}

Simultaneous observations in the SiO J=1-0 v=1 and 2 transitions 
at 43.122 and 42.821 GHz, respectively, were made 
with the 45-m radio telescope at Nobeyama 
on April 25, May 10, 11, and 12, 1997. 
A cooled SIS receiver with a band width of 
about 0.4 GHz was used and the system temperature 
(including atmospheric noise) was 190-200 K.
The aperture efficiency of the telescope was about 0.60 at 43 GHz.
The half-power beam width (HPBW) was about 40$''$ at 43 GHz. 
A factor, 3.6 Jy/K, was used to
convert antenna temperature to flux density.  
An acousto-optical spectrometer array of a low-resolution (AOS-W) 
was used. Each spectrometer has a 250 MHz band width and 2048 frequency 
channels, giving the velocity coverage of about 1700 \kmps and 
the spectral resolution of 1.7 \kmps (per two binned channels).
Observations were made in a position 
switching mode and the off position was chosen 10$'$ away from 
the Galactic center in azimuth.  

We observed three positions near the Galactic 
center: one at the position of Sgr A*,
(RA, Dec, Epoch)=(17h42m29.314s, -28\degree59$'$ 18.3$''$ , 1950)
[(l,b)=(359.944\degree, -0.046\degree); \cite{rog94}], and the other
two at the positions offset by one beam on either side 
of Sgr A* in the Galactic longitude, 
($\Delta$l,$\Delta$b)=($\pm$40$''$,0$''$).
We performed the observations on clear nights without wind. 
Telescope pointing was carefully checked 
before and after each observation 
 using a nearby strong SiO maser source, 
OH2.6-0.4. The average pointing accuracy was confirmed to be 
 better than 5$''$. The total integration time was 6.8 h 
 for the center and
3.0 h each for the offset positions.
For the center, ($\Delta$l,$\Delta$b)=(0$''$,0$''$), 
we combined the spectra taken on April 25 and May 10, 1997. 
For the offset positions, data were taken on  
May 11 and May 12. Weather condition on May 12 was not perfect
when the position  (+40$''$,0$''$) was observed; 
$T_{sys}$ was about 220 K and the wind speed about 5 m s$^{-1}$  
(which might have caused the pointing error of about 5$''$).
Thus, the noise level on the spectrum at (+40$''$,0$''$)
was slightly higher than that on spectra taken on the other days.

Obtained spectra toward Sgr A* exhibited 
a baseline offset of about 3.2 K 
from zero due to the continuum emission. This corresponds to a 
flux density  of about 12 Jy at 43 GHz, 
which is consistent with the
previous measurement of the Sgr A*(11.6 Jy) at 43 GHz 
with the 45-m system (\cite{sof86}).
The spectra exhibited a baseline distortion 
(of about 0.3 K at the maximum) and weak ripples 
probably due to standing waves in the 45-m telescope system.
The ripples in the velocity range of
$\pm 350$ \kmps of the SiO lines
were relatively weak (of about 0.02 K on average).
In order to remove complex ripple features from the spectra,
we have taken running means of the spectra 
(average of 100 channels, or about the 80 km/s width) and 
 averaged spectra were subtracted from originals. 
With this procedure,  the baseline of the
resulting spectra became quite flat. Because
the SiO maser lines are quite narrow (the widths less than 10 km/s) 
and weak (T$_{a} < 0.2$ K), this method seems to work well. 
There are other molecular and atomic lines possibly contaminating  
the SiO J=1-0 v=2 line at 42.821 GHz: $^{29}$SiO J=1-0, v=0 
(42.879 GHz), H53$\alpha$ (42.891 GHz),
SiC$_{4}$ J=14-13 (42.945 GHz), and C$_{6}$H J=31/2-29/2 
(42.970 and 42.970 GHz). 
These lines might appear as broad features at the Galactic center,
and may overlap somewhat with the SiO J=1-0 v=2 line,
causing a bad baseline. 
With this baseline removal, broadline features 
of the above mentioned molecules, even if present, were eliminated.
The spectra finally obtained are shown in figure 1.
The rms noise temperatures at the center,
($\Delta$l,$\Delta$b)=(0$''$,0$''$),  are 0.004 and 0.009 K 
at 42.821 GHz (SiO J=1-0, v=2) and 43.122 GHz (SiO J=1-0, v=1), 
respectively. The spectrometer used for the v=1 line gave 
systematically higher rms value of noise. This is probably due to a
calibration error of the photo-diode levels in 
the spectrometer.

\placefigure{fig1}

Detections of the SiO maser components were judged 
by the following criteria (both must be satisfied).
\begin{itemize}
\item{Line intensities in either transition must be 
greater than 5 times 
the RMS noise level with line widths greater than three channels (2.5 
\kmps).}
\item{Emission peaks are seen at the same velocity 
 in both the J=1-0, v=1 and 2 transitions (within 2 \kmps ), 
 or at least,  a peak is seen 
in one transition and a positive deflection (not trough)
is seen in the other transition.}  
\end{itemize}
These criteria worked nicely 
in our experiences, guarding against identification
as signal of anomalous noises known in AOS data. 
For the case of a short integration,
the second criterion has been quite useful and follow-up
integrations revealed the SiO lines very often.

SiO maser components detected at the three positions are shown 
in table 1. It is well known that the SiO maser velocity of a single
source coincides  with the stellar velocity, i.e.,  
with the middle of the OH 1612 MHz double-peak velocities within a few 
\kmps (\cite{jew91}; \cite{jia95}).
Therefore, if the velocity of the detected SiO feature 
is close to the velocity 
of an already known OH 1612 MHz source at the Galactic center, 
it is highly 
probable that the SiO source is associated with the OH source. 
With this velocity coincidence, we have assigned probable
associations with OH/IR sources in table 1. 

\placetable{tb1}

\section{Discussion}

The radial velocity of SiO maser emission indicates 
the stellar velocity, i.e., the velocity of a central star. 
Velocity widths  
are known to be less than 10 \kmps in most of SiO maser sources. 
Especially for the case of a source at the Galactic center, 
we can detect only a narrow emission peak and not a weak broad pedestal. 
From these facts,  we can safely assume that the detected SiO peaks,  
if separated more than 10 \kmps, come
from individual sources in the telescope beam. In the following 
discussion, we regard each SiO component listed in table 1 
as an individual source. 

\subsection{Associations with previously known objects at the 
Galactic center}  

Extensive surveys of the Galactic-center region have been made 
before with the VLA in OH 1612 MHz (\cite{lin92b}; \cite{sjo96})
and in H$_{2}$O and SiO lines (\cite{lev95}; \cite{men97}). 
Because sources detected in this paper are within a 1$'$ radius from
the Galactic center,  4 sources in the list of Lindqvist et 
al. (1992b) are considered to be candidates for associations; 
OH359.939-0.052 ($V_{av}=52.2$ \kmps), 
OH359.946-0.047 ($V_{av}=-27.2$ \kmps), 
OH359.952-0.036	($V_{av}=82.2$ \kmps), and 
OH359.954-0.041	($V_{av}=70.2$ \kmps). 
In addition, we take into account 5 SiO/H$_{2}$O sources listed 
by Menten et al. (1997), one of which coincides 
with OH359.946-0.047. 

The judgment for association is based on the velocity 
coincidence. The position ($\pm$ 20$''$) and velocity
($\pm$ 3 \kmps) coincidences are shown as footnotes in table 1 
for each SiO components. The radial velocities of the source 
No. 3 (-25.2 \kmps) at (0$''$, 0$''$) 
and the source No. 3 ($V_{lsr}=-24.5$ and $-28.3$ \kmps) at 
(+40$''$,0$''$), 
are quite close.  It is possible that they are the same source
that is located at the middle of the two beams. 
However, the source No. 3 at (0$''$,0$''$) 
is probably associated with SiO(+7.7,+4.2) =
IRS 10 EE in table 1 of Menten et al. (1997) and
OH359.946-0.047 in Lindqvist et al. (1992b). 
In this case, the intensity of this source at (+40$''$, 0$''$) should be
7 \% of the signal intensity at (0$''$, 0$''$),
or at most  14 \% even if a pointing error of 
5 $''$ is taken into account,
indicating that it is undetectable at (+40$''$, 0$''$). Therefore,  
we conclude these two emissions at similar velocities 
come from different sources.
  
\subsection{Velocity distribution}
It is interesting to examine how the average velocity of sources 
varies with Galactic longitude, because fast rotating gas components
of about 200-300 km/s have been found in the Galactic center spiral 
of an angular size of about 30$''$ (\cite{loc83}; \cite{lac91}). 
We have made a linear regression analysis of $V_{lsr}$ with $l$. 
Source positions were assumed to be exactly at beam centers, i.e., 
 -40$''$, 0, and +40$''$ in longitude.
The least-square fit to the radial velocities of the detected sources gives 
$V_{lsr} = -25.5 [\pm 32.2] -0.098[\pm 1.14] (\Delta l/'') $  \kmps, 
indicating a quite small velocity gradient. 

  It can be considered that the high-velocity source No. 7 
at (0$''$, 0$''$) may influence the results. 
Because the SiO intensity of this source is quite strong, this 
high velocity source may be located closer to the Sun than
 the other sources on the same line of sight.
We have also made the linear fit excluding this 
source and obtained
$V_{lsr} = -0.9 [\pm 21.6] -0.010[\pm 0.734] (\Delta l/'') $  \kmps. 
The result does not differ significantly from the previous one.

The velocity dispersion (deviation from the regression line) 
is 111.4 \kmps for the set of all sources and
71.2 \kmps for the set excluding the high velocity source No. 7.
This value is comparable with the velocity dispersion of
$\sim$125 \kmps of infrared stars within 20$''$ from the center
(\cite{mcg89}). However, it seems significantly smaller than
154 ($\pm$ 19) \kmps for the Helium-rich blue supergiants/Wolf-Rayet 
stars with distances of 1-12$''$ from the Galactic center (\cite{kra95}).
 
Lindqvist et al. (1992a) found that
the linear regression of the (l-v) diagram for their 134 OH/IR sources 
is about $V_{lsr}= -7 + 180 \: (\Delta l/deg) $ \kmps, or if expressed 
in the above unit, $V_{lsr}= -7 + 0.05 \: (\Delta l/'')$ \kmps. 
 The gradient is as small as the value
obtained in this paper. Lindqvist et al. (1992a) mentioned that 
the smaller number of samples within 10 pc (about 4$'$) 
from the Galactic center gave the larger velocity gradient.
From their figure 9, we can estimate this gradient for the smaller 
sample as 0.83 \kmps /$''$. McGinn et al. (1989) also gave
the velocity gradient of $\sim$1.0 \kmps /$''$ 
from infrared stars within 100$''$ from the center 
(estimated from figure 4 in their paper).
These values are consistent
 with the present non-detection of rotational motion within the errors. 
Genzel et al. (1996) found that late-type stars 
within 5$''$ from the center exhibit a gradient of about 5 
\kmps /$''$, which is somewhat larger than the present result.

If SiO sources near the Galactic 
center have the same kinematic properties as the stars 
detected by McGinn et al. (1989) and Lindqvist et al. (1992b), i.e., 
they belong to the nuclear disk of a size of more than a few arc minutes,
the nondetection of the rotational motion in the present paper
seems reasonable. The velocity dispersion obtained 
is also consistent
with the previous measurements for the stellar disk.

\subsection{Source number density} 

The source numbers we detected are 4, 7, and 3 at the three positions,
(-40$''$,0$''$), (0$''$, 0$''$), and (+40$''$, 0$''$), respectively. 
The apparent decrease in 
the source number at the off-center positions is probably due to the 
larger noise level (less integration time) at these positions.
In fact, if we correct the source counts for the noise levels, assuming 
that we try to detect the (0$''$, 0$''$) No. 1--7 sources 
with higher noise levels,
the expected detections above 5 sigma are 3 and 1 
with the noise levels at 
(-40$''$,0$''$) and (+40$''$, 0$''$), respectively. 
Taking into account statistical uncertainties,
we conclude that the source surface number density 
is not peaked at the Galactic center on a scale of about 40$''$.

It is interesting to compare the SiO source surface density 
with the density of OH/IR sources. 
Lindqvist et al. (1992a) gave the OH 1612 MHz source density of
1700 per square degree, which corresponds to about 0.47 
sources per square arcminute. Apparently the number of SiO sources 
at (0$''$, 0$''$)
(=7 per 40 \arcsec \ beam) is about 40 times 
that expected from the OH/IR survey. The number density of 
SiO maser sources found in the present survey is an order of magnitude  
larger than the density of  OH 1612 MHz maser sources 
which were detectable with the VLA. 
Hence we would like to call
the cluster of SiO maser lines at the Galactic center 
a ``SiO maser forest''. 

OH/IR stars with OH 1612 MHz emission are highly obscured red 
supergiants, while SiO maser stars are very often less 
massive semiregular or Mira variables which have weak 
or no OH 1612 MHz emission. For example, 
W Hya, a semiregular variable at the distance of 95 pc 
(\cite{rei97}), exhibits strong SiO maser emission of about
700 Jy in the J=1-0 v=1 and 2 transitions, 
but shows no OH 1612 MHz line. 
If it is placed at the Galactic center (at the distance of 8 kpc),
the flux density of the same line would be 0.1 Jy, which is 
comparable with the flux densities observed in the present survey.
Therefore, many of the SiO sources we have detected 
at the Galactic center
might be Mira and semiregular variables. 
While the statistics of a sample of 172 bulge OH/SiO sources 
(observed by both OH and SiO lines;  \cite{jia95}) shows 
that the number of SiO sources without OH is comparable
with the number of OH sources with and without SiO. If the same
statistics applies to the Galactic center sources, we
should detect more OH 1612 MHz sources.
Therefore, some violent mechanism 
might strip the outer (OH emitting) envelope of AGB stars
near the Galactic center, causing the underdensity of 
OH 1612 MHz sources at the Galactic center.

\section{Conclusions}
 
   We have detected 14 SiO maser sources at the Galactic center 
and at the 40$''$ off-center positions; 6 of these sources are
associated with already known OH 1612 MHz or SiO sources, and
8 are new detections.
\footnote{Menten and Reid (1997) recently reported 
 the detections of several more SiO sources [two of them are
 associated with No. 5 and 7 at (0$''$, 0$''$) in Table 1].}
We find that the radial velocity gradient
with respect to Galactic longitude is small. 
The SiO source number density is an order of magnitude 
higher than the density of previously found OH 1612 MHz objects 
toward the Galactic center. The flat distribution of the source
number density around the center and the velocity dispersion 
indicate that these SiO sources are associated with the Galactic 
nuclear stellar disk. The small radial velocity gradient of the 
sources also does not conflict with this interpretation. 

\acknowledgments

We thank David Morris for reading the manuscript. We also thank 
Drs. N. Ukita,  Y. Nakada,  O. Kameya, S. Matsumoto, and H. Imai
for discussions and encouragements. 
This research was partly supported by Scientific Research 
Grant (C) 08640337 from the Ministry of Education, 
Science, Sports, and Culture, Japan.


\clearpage
\figcaption[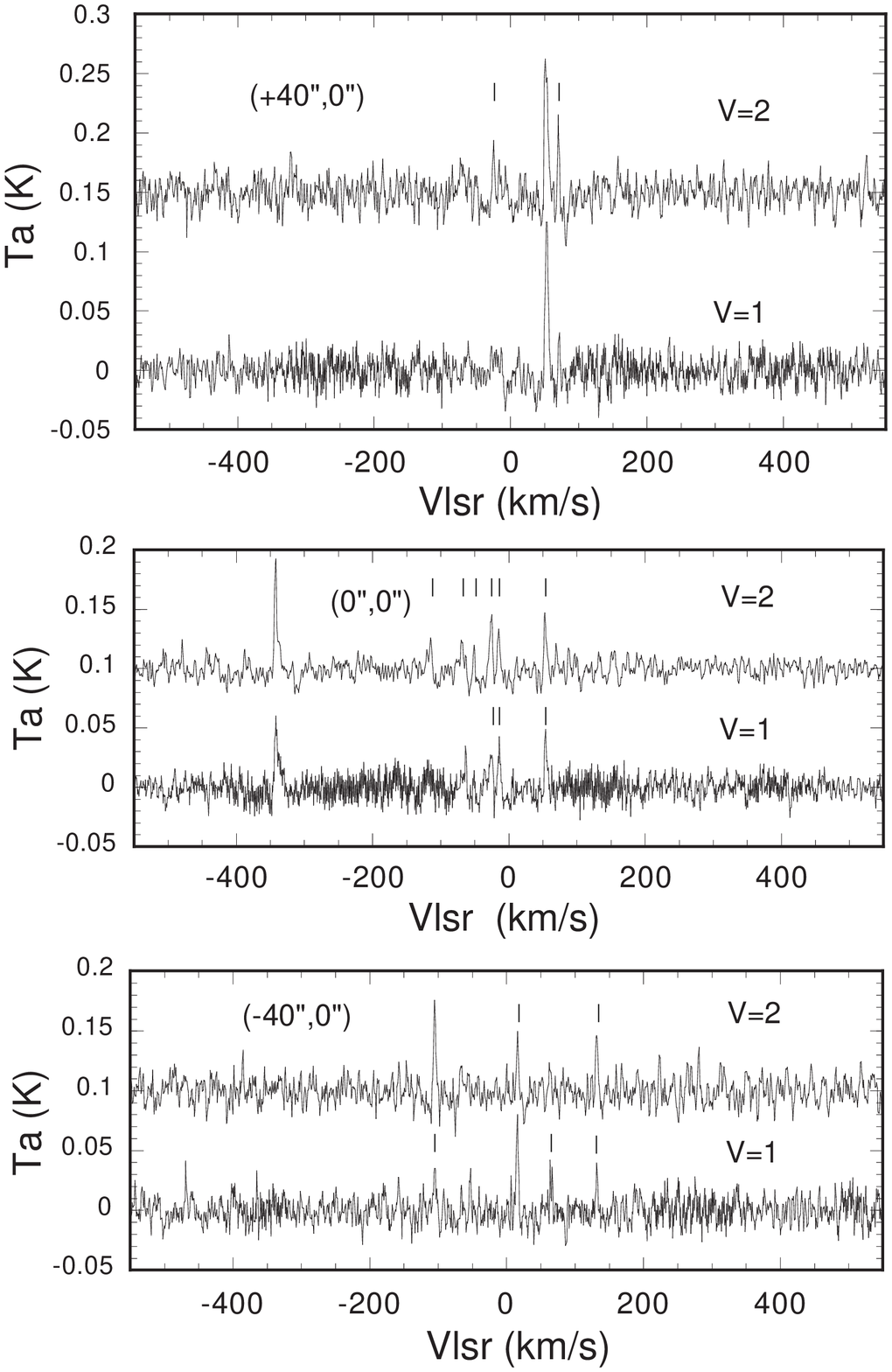]{Spectra of the SiO J=1-0 v=1 and v=2 
transitions toward the Galactic center; top at ($\Delta$ l,$\Delta$ b)=
(+40$''$, 0$''$), middle at (0$''$,0$''$), and bottom at (-40$''$,0$''$). 
Detections except strongest ones are marked by ticks.
\label{fig1}}

\begin{deluxetable}{clccccccccc}
\tablenum{1}
\tablewidth{0pt}
\tablecaption{List of detected SiO peaks (sources) \label{tb1}}
\tablehead{  & &\colhead{v=2}  
&\colhead{J=1-0} &  &\colhead{v=1}  & \colhead{J=1-0}&  \\
 \cline{3-8} 
  \colhead{SiO l,b (\degree)} & \colhead{No}
& \colhead{\Vlsr}  & \colhead{I$_{peak}$} & \colhead{F}
& \colhead{\Vlsr}  & \colhead{I$_{peak}$}  & \colhead{F} & 
 \\
 ($\Delta l,\Delta$ b)  &
  &(\kmps) &(Jy) &(Jy\kmps)
  &(\kmps )&(Jy) &(Jy\kmps)
    }

\startdata
SiO359.933-0.046 &
1&131.2& 0.167& 0.647& 131.2& 0.143& 0.380 \\
(-40'', 0'')&2&---& --- & --- & 63.7& 0.152& 0.227 \\
&3\tablenotemark{a}&15.9& 0.179& 0.497& 15.9& 0.289& 0.942 \\
&4&-105.1& 0.273& 0.921& -104.2& 0.127& 0.511 \\

SiO359.944-0.046 &					
1&53.1&	0.169& 0.712& 54.4& 0.177& 0.738 \\
 (0'', 0'')&2\tablenotemark{b}&-14.6& 0.121& 0.435& -13.8& 0.153& 0.440\\
&3\tablenotemark{c}&-25.2& 0.164& 0.797& -25.2& 0.099& 0.560\\
&4&-51.2&0.072&0.143&---&---&---\\
&5&-69.6&0.087&0.575&---&---&---\\
&6\tablenotemark{d}&-114.9& 0.094& 0.497&---&---&---\\
&7&-341.9& 0.333& 0.195& -341.9& 0.216& 1.395\\

SiO359.955-0.046 &					
1\tablenotemark{e}&70.0& 0.234& 0.494& --- & --- & ---\\
(+40'', 0'')&2\tablenotemark{f}&50.6& 0.403& 2.377& 52.6& 0.450& 2.173\\
&3&-24.7& 0.158& 0.474& --- & --- & ---\\
\enddata
\footnotesize
\tablenotetext{a}{=OH359.931-0.050, V$_{av}$=17 \kmps
  (\cite{sjo97})}
\tablenotetext{b}{=IRS15NE =SiO(+1.2,+11.3), V$_{av}$=-15 \kmps (\cite{men97})}
\tablenotetext{c}{=OH359.946-0.047=IRS10 EE,  V$_{av}$=-27 \kmps 
(\cite{lin92b})}
\tablenotetext{d}{=IRS7=SiO(0.0,+5.6), V$_{av}$=-120 \kmps (\cite{men97})}
\tablenotetext{e}{=OH359.954-0.041, V$_{av}$=70 \kmps (\cite{lin92b}) }
\tablenotetext{f}{=OH359.956-0.050, V$_{av}$=50 \kmps
  (\cite{sjo96}; \cite{lev95})}
\end{deluxetable}

%

\begin{thebibliography}{}

\bibitem[Blum et al.\ 1996]{blu96}Blum, R. D., Sellgren, K.,
 \& DePoy, D. L. 1996, ApJ, 470, 864


\bibitem[Eckart \& Genzel 1997]{eck97}Eckart, A., \& Genzel, R. 
1997, MNRAS, 284, 576

\bibitem[Eckart et al.\ 1993]{eck93}Eckart, A., Genzel, R., Hofman, 
R., Sams, B. J., \& Tacconi-Garman, L. E. 1993, ApJ, 407, L77

\bibitem[Genzel et al.\ 1996]{gen96}Genzel, R., Thatte, N., Krabbe, A., 
Kroker, H., \& Tacconi-Garman, L. E. 1996, ApJ, 473, 153
   

\bibitem[Izumiura et al.\ 1995]{izu95}Izumiura, H., Deguchi, S., 
     Hashimoto, O., Nakada, Y., Onaka, T., 
     Ono, T., Ukita, N.,  \& Yamamura, I. 1995, 
     ApJ,   453,  837 
     
\bibitem[Jewell et al.\ 1991]{jew91}Jewell, P. R., Snyder, L. E., 
     Walmsley, C. M., Wilson, T. L., 
   \& Gensheimer, P. D. 1991,  A\&A 242, 211
   
\bibitem[Jiang et al.\ 1995]{jia95}Jiang, B. W., Deguchi, S., 
      Izumiura, Y., Nakada, Y., \& Yamamura, I.  1995, 
        PASJ  47, 815

\bibitem[Krabbe et al.\ 1995]{kra95}Krabbe, A,, Genzel, R., Eckart, A., 
Najarro, F., Lutz, D., Cameron, M., Kroker, H., Tacconi-Garman, L. E.,
Thattle, N., Weitzel, L., Draper, S., Geball, T., Stergnberg, A., \& 
Kudritzki, R. 1995, ApJ, 447, L95

\bibitem[Lacy et al.\ 1991]{lac91}Lacy, J.ÊH., Achtermann,ÊJ.ÊM., 
 \& Serabyn,ÊE. 1991, ApJ, 380, L71
 
\bibitem[Levine et al.\ 1995]{lev95}Levine, D., Figer, D. F., Morris, M.,
\& Mclean, I. S., 1995, ApJ, 447, L101

\bibitem[Lindqvist et al.\ 1991]{lin91}Lindqvist, M., 
Winnberg, A., Johansson,ÊL.ÊE.ÊB., \& Ukita, N. 1991, A\&A, 250, 431

\bibitem[Lindqvist et al.\ 1992a]{lin92a}Lindqvist, M., 
  Habing, H. J., \& Winnberg, A. 1992a, A\&A, 259, 118
  
\bibitem[Lindqvist et al.\ 1992b]{lin92b}Lindqvist, M., Winnberg, A.,
  Habing, H. J., \& Matthews, H. E.1992b, A\&AS, 92, 43  
  
\bibitem[Lo and Claussen\ 1983]{loc83}Lo, K. Y., \& Claussen, M. J.
 1983, Nature 306, 647

\bibitem[McGinn et al.\ 1989]{mcg89}McGinn, M. T., Sellgren, K.,
Becklin, E. E., \& Hall, D. N. B. 1989, ApJ, 338, 824  

\bibitem[Menten \& Reid \ 1997]{mar97}Menten, K., \& Reid, M. 
 1997, a poster paper presented 
in IAU Sympo. 184. held Aug 18-21, 1997, in Kyoto, Japan.

\bibitem[Menten et al.\ 1997]{men97}Menten, K., Reid, M., Eckart, A., 
\& Genzel, R. 1997, ApJ, 475, L111 

\bibitem[Morris \& Serabyn\ 1996]{mor96}Morris, M., \& Serabyn, E. 
1996, ARAA, 34, 645

\bibitem[Reid \& Menten\ 1997]{rei97}Reid, M. J., \& Menten, K. M. 
1997, ApJ, 476, 327
   
\bibitem[Rogers et al.\ 1994]{rog94}Rogers, A. E. E., et al. 1994, 
ApJ, 434, L59

\bibitem[Shiki et al.\ 1997]{shi97}Shiki, S., Ohishi, M., 
\& Deguchi, S. 1997, ApJ,  478, 206


\bibitem[Sjouwerman \& van Langevelde\ 1996]{sjo96}Sjouwerman, L. O., 
\& van Langevelde, H. J. 1996, ApJ, 461, L41 

\bibitem[Sjouwerman et al.\ 1997]{sjo97}Sjouwerman, L. O., 
van Langevelde, H. J., Winnberg, A., \& Habing, H. J. 1997, 
A\&A, in press


\bibitem[Sofue et al.\ 1986]{sof86}Sofue, Y., Inoue, M.,
Handa, T., Tsuboi, M., Hirabayashi, H., Morimoto, M., \& Akabane, K. 
1986, PASJ, 38, 475


\bibitem[Winnberg 1996]{win96} Winnberg, A. 1996, 
``The Galactic Center'', ASP conf. ser. Vol.102, p294, 
ed. by Gredel et al. (\it{Astron. Soc. Pacific; San Francisco})

\end{thebibliography}
\end{document}